# Infrared laser stimulated broadband white emission of transparent Cr:YAG ceramics obtained by solid state reaction sintering

M.A. Chaika, R. Tomala, W. Strek

*Institute of Low Temperature and Structure Research Polish Academy of Science, Okolna 2, 50-422 Wroclaw, Poland.*

**Abstract**

In the present work, the Light Induced White Emission (LIWE) was investigated as one of the sources of laser loss in transparent Cr:YAG ceramics. We have found that transparent Cr:YAG ceramics is capable of generating bright LIWE at excitation powers above a certain threshold. Intensity of LIWE strongly depends on the excitation power and ambient pressure. The host temperature estimated from $Cr^{3+}$ luminescence was found to be below 600°C, while the most intense white emission was found between 50-400°C. The mechanism of the laser induced white emission was discussed in terms of Intervalence Charge Transfer (IVCT) in chromium mixed valence pair.

*Keywords:* Ceramics; Cr:YAG; Tetravalent chromium; White light emission; Intervalence Charge Transfer

**Introductions**

Recently, the intense Laser Induced White Emission (LIWE) has been investigated in numerous studies under excitation with a focused infrared laser beam [1-8]. LIWE has been recently observed by us in Yb:YAG [1], $Er:Sr_2CeO_4$ [2], $Nd:Sr_2CeO_4$ [3], Yb:GdAG [4], $Nd:Y_2Si_2O_7$ [5], $LiYbF_4$ [6], $NdAlO_3$ [7], and graphene nanoparticles [8]. The broadband white emission can be achieved from the samples placed in vacuum ($10^{-2}$ mbar) excited by focused near infrared laser beam [2]. LIWE is a threshold phenomenon that strongly depends on the excitation power density [5], number of photons involved in upconversion process [4], rise and decay times [3]. LIWE possesses a high efficiency which is close to 10% [1] and relatively low temperature of the matrix during broadband white emission [6]. This phenomenon can be useful for creating new sun-like emission sources [1-3], but it can also be a source of loss processes in laser materials [7,9]. Until now, most papers studied LIWE in nanopowders or translucent nanoceramics, while the study of LIWE in transparent laser materials remained a gap in the research.

$Cr^{4+}$:YAG has received a lot of attention due to its potential use as a passive Q-switches for solid state laser [10,11]. The YAG formula can be written as [C3][A2][D3]O12, where C, A, and D denote cation sites which are coordinated by oxygen atoms in a dodecahedral, octahedral, and tetrahedral positions, respectively [12-14]. The A and D sites are occupied by Al ions, whereas C sites are taken by Y ions. An interesting feature of $Cr^{4+}$ ions in YAG matrix is the ability to occupy two crystallographic positions and the need to use divalent impurities such as $Ca^{2+}$ or $Mg^{2+}$ as charge compensators for $Cr^{4+}$ ions [10].

The tetrahedrally coordinated $Cr^{4+}$ ions are responsible for laser properties of $Cr^{4+}$:YAG, so various energy loss processes for these ions can reduce the performance of $Cr^{4+}$:YAG based laser. The possibility of using $Cr^{4+}$:YAG materials as Q-switched lasers as well as tunable lasers is based on the large ground-state absorption cross section and excited-state lifetime of tetrahedrally coordinated $Cr^{4+}$ ions allowing to reach considerable depletion of the ground state [15]. The lasing efficiency of high-quality $Cr^{4+}$:YAG crystals does not extend 10% indicating a significant energy loss in this material [16]. The exited state absorption is the one of the known parasitic processes in $Cr^{4+}$:YAG materials [15]. LIWE, which is based on multiphoton absorption, is known as an energy loss process in laser materials [7]. As far as we know, there are still no articles reporting LIWE in transparent Cr:YAG ceramics.

In the present work, LIWE induced by infrared laser excitation was studied in transparent Cr:YAG ceramics and the model for describing LIWE is proposed.

**Experimental**

The Cr:YAG samples were sintered in vacuum furnace by solid state reaction. High purity $Al_2O_3$ (purity >99.99%, Baikowski, d=0.15-0.3 μm), $Y_2O_3$ (purity >99.999%, Alfa Aesar, d=<10 μm), $Cr_2O_3$ (purity >99.97%, Alfa Aesar, d=<100 nm), CaO (purity >99.999%, Sigma Aldrich, d=<0.1 μm) were used as starting materials. Powders were taken in stoichiometric ratio, concentrations of Ca and Cr were taken in order to replace Y and Al, respectively. $Cr_2O_3$ and CaO powders were weighted precisely to obtain chromium and calcium content of 0.1 and 0.5 at.%, respectively [10,11]. Homogenization was performed by ball milling for 15 hours using high purity $Al_2O_3$ balls. The milled slurry was dried for 1 day in air and sieved through a 200-mesh screen. The compacts were prepared by applying isostatic pressing at P = 250 MPa. Sintering was performed at 1750°C for 50 hours using solid state reaction (SSR) in vacuum furnace.

Before measurements the samples were polished with diamond abrasive with a gradual decrease in the size of the abrasive from 30 to 7 μm. Fig. 1 presents photograph of the sample. After processing, the tablets of 8 mm in diameter and 1 mm thick were obtained.

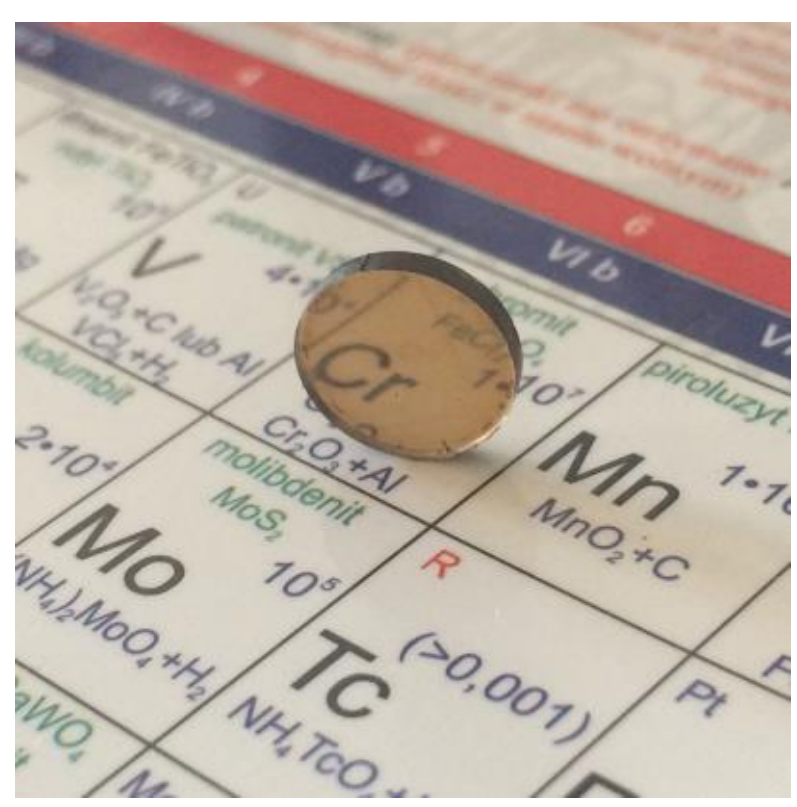

Fig. 1. Photo of the transparent Cr:YAG ceramics after vacuum sintering and following air annealing.

Optical spectroscopy, TEM and X-ray measurements were used for characterization of the samples. The X-ray diffraction spectra were measured using X'Pert PRO powder diffractometer (PANalytical, The Netherlands) equipment with a linear PIXcel detector under Cu Kα radiation (λ = 1.54056 A). The ceramics was examined using Titan$^3$ Transmissions Electron Microscopy. LIWE was excited by a focused Nd:YAG laser (1064 nm) with a cross section up to 0.18 mm, and maximum laser output of 3.4 W, measurements were done in vacuum. Due to reflections of the laser beam at the surfaces of the lens and the vacuum chamber, the power of laser beam at sample surface was reduced by a factor of 0.88. Therefore, the maximum laser output on the surface of the sample was 3.0 W, which is makes it possible to achieve a laser density up to $9.3 \cdot 10^3$ W/cm$^2$.

The samples were placed into a vacuum cell connected to an EXT75DX Turbo Molecular High Vacuum Pump with an integrated TIC controller (Edwards) allowing measurements at a low pressure of $10^{-4}$ mbar. The emission spectra were measured using CW Nd:YAG laser (CNI Optoelectronics Tech. Co., Ltd.) with λ = 1064 nm, beam divergence >3 mrad, and power stability <0.2% as an excitation source. AVS-USB2000 Avantes Spectrometer was used for collection of emission spectra.

## Results

Before the experiments on white emission, the homogeneity of the samples was evaluated. Fig. 2 show the XRD pattern of Cr:YAG ceramics. The X-ray diffraction study has shown the absence of any impurity phases in the samples. Diffraction data were refined with the cubic Ia-3d space group of $Y_3Al_5O_{12}$ using the Rietveld analysis [17]. The analysis gives a lattice parameter of 12.0112±0.0007, all diffraction peaks can be well indexed as pure YAG phase, and no impurity phases are observed.

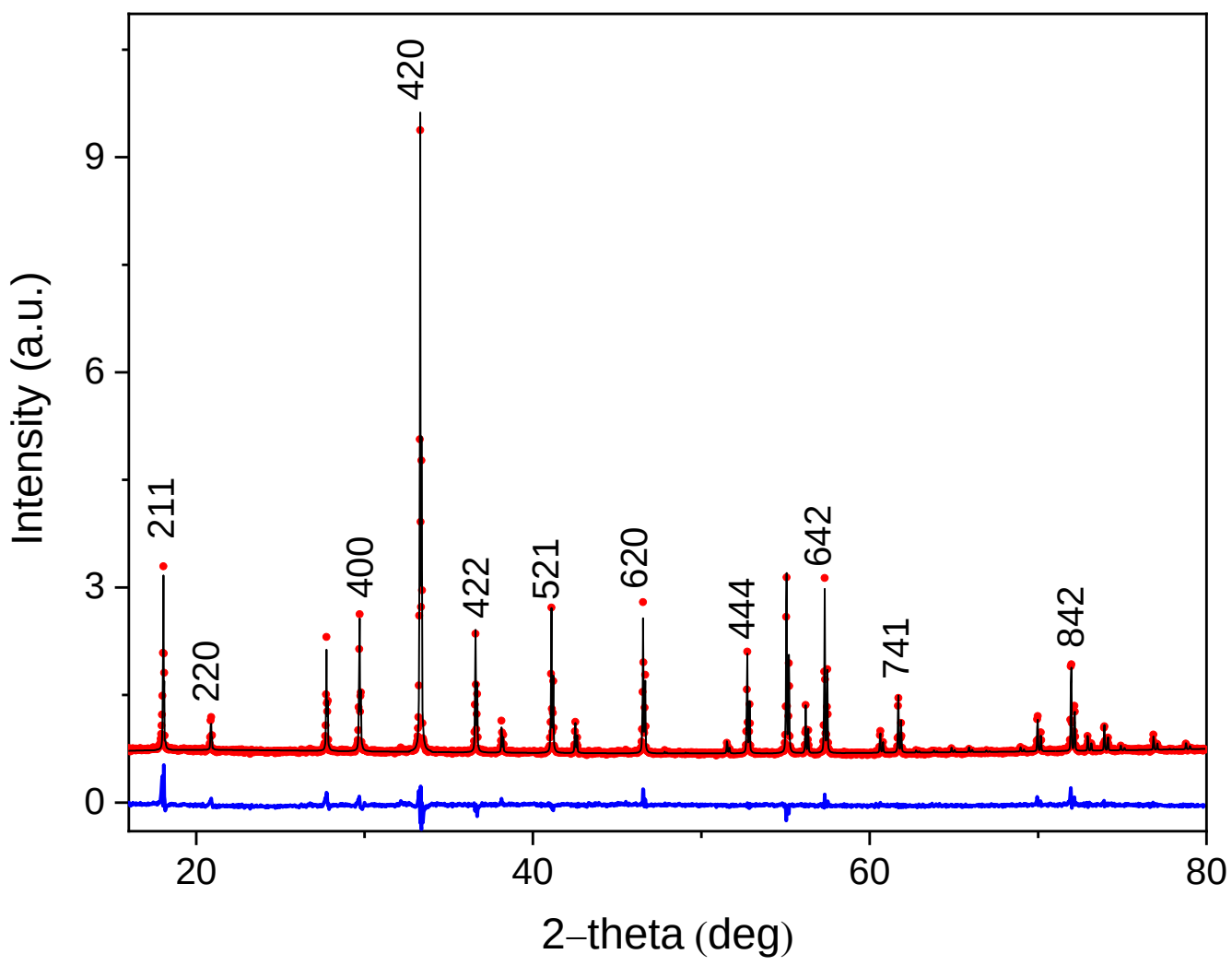

Fig. 2. X-ray diffraction pattern of Cr:YAG ceramics. Insets: results of Rietveld refinement analysis and the grain size distribution, respectively.

TEM didn't show the presence of any impurity phases with the exception of small fraction of $Al_2O_3$. This impurity phase is located between the grains which size is up to 0.5 μm. The EDX analysis has detected the presence of Y, Al and O as shown in Fig. S1. The grain boundary is <1 nm thick with lattice fringes extending to the boundary of each grain (Fig. S2). The grain boundaries and triple lines with no evidence of crystalline or amorphous second phases. The grain sizes were in the range from 0.5 to 10 μm with the average grain size of 2.0±0.3 μm (inset in the Fig 3). The average grain size was determined by the linear intercept method; The Students t-test was applied to calculate the standard deviation [10].

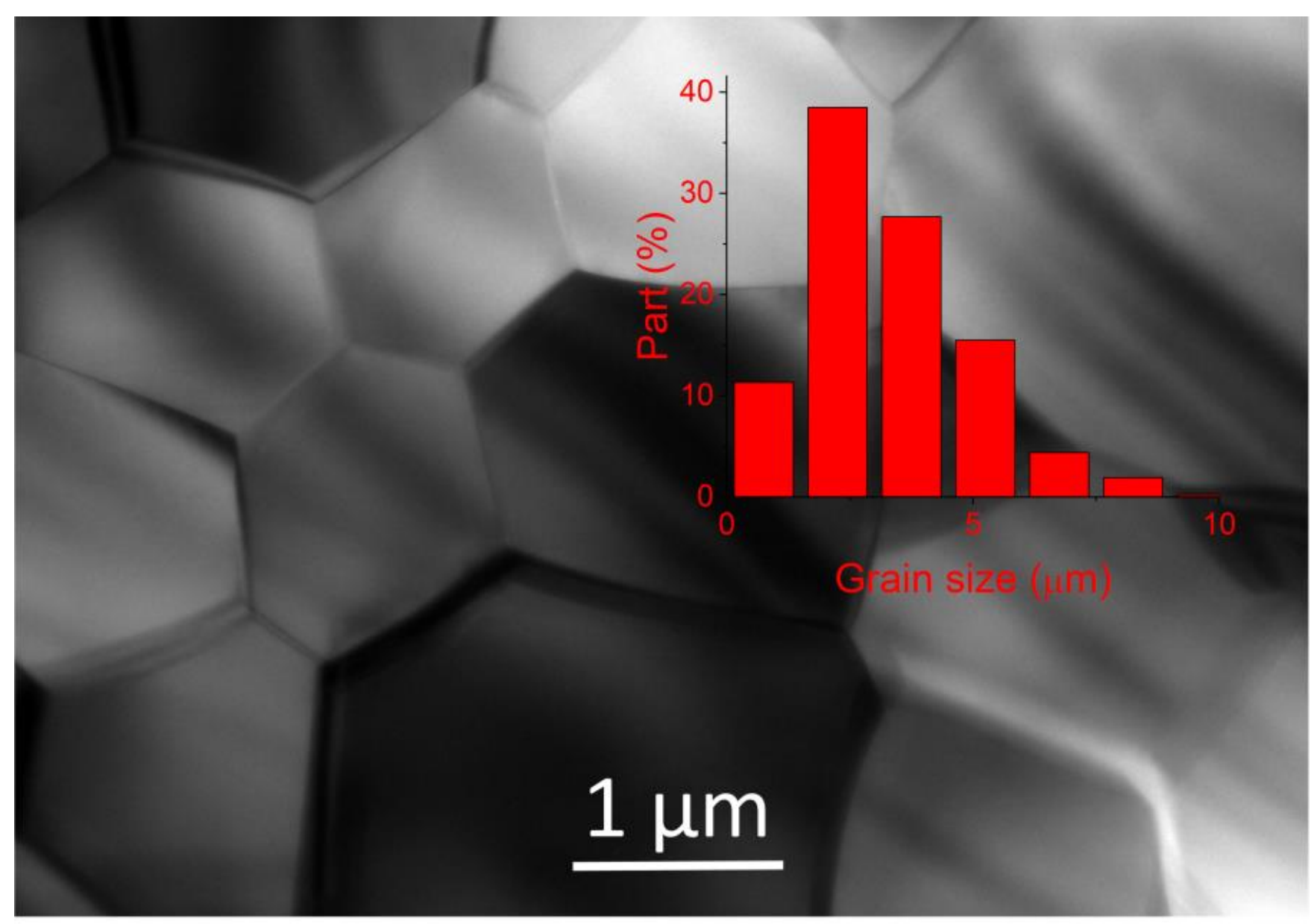

Fig. 3: STEM image of a Cr:YAG ceramics after thermal etching. Inset: Grain size distribution of Cr:YAG ceramics.

The optical measurements have shown that up to 20% of the total amount of chromium ions is in tetravalent state. The transmission spectra of Cr:YAG ceramics are shown in Fig. 4. The thickness of the samples was about 1 mm. After vacuum sintering, the samples contain only Cr in trivalent state. Formation of $Cr^{4+}$ ions in Cr:YAG ceramics occurs during annealing in an oxidizing ambient, where tetravalent impurities can be stabilized by the formation of structural defects in the lattice [11]. After annealing in air at 1450°C for 20 h, the ceramics colour changed from light green, typical for $Cr^{3+}$ to dark-brown typical to $Cr^{4+}$ included into YAG. Absorption spectrum of $Cr^{4+}$ ions in $Cr^{4+}$:YAG ceramics consist of the bands corresponding to octahedral $Cr^{4+}$ (1-4), tetrahedral $Cr^{4+}$ (5-8) and octahedral $Cr^{3+}$ ions (9-10) (Fig. S3) [10]. Concentration of octahedral and tetrahedral $Cr^{4+}$ was $2.9{\cdot}10^{18}cm^{-3}$ and $1.9\ 10^{18}cm^{-3}$, respectively (more details see in [18]). In calculate to $Al^{3+}$ ions, the 0.1 at.% of Cr was used in the samples which are $23\ 10^{18}{\cdot}cm^{-3}$. Therefore, 1/5 of the total number of Cr ions are the tetravalent state after air annealing.

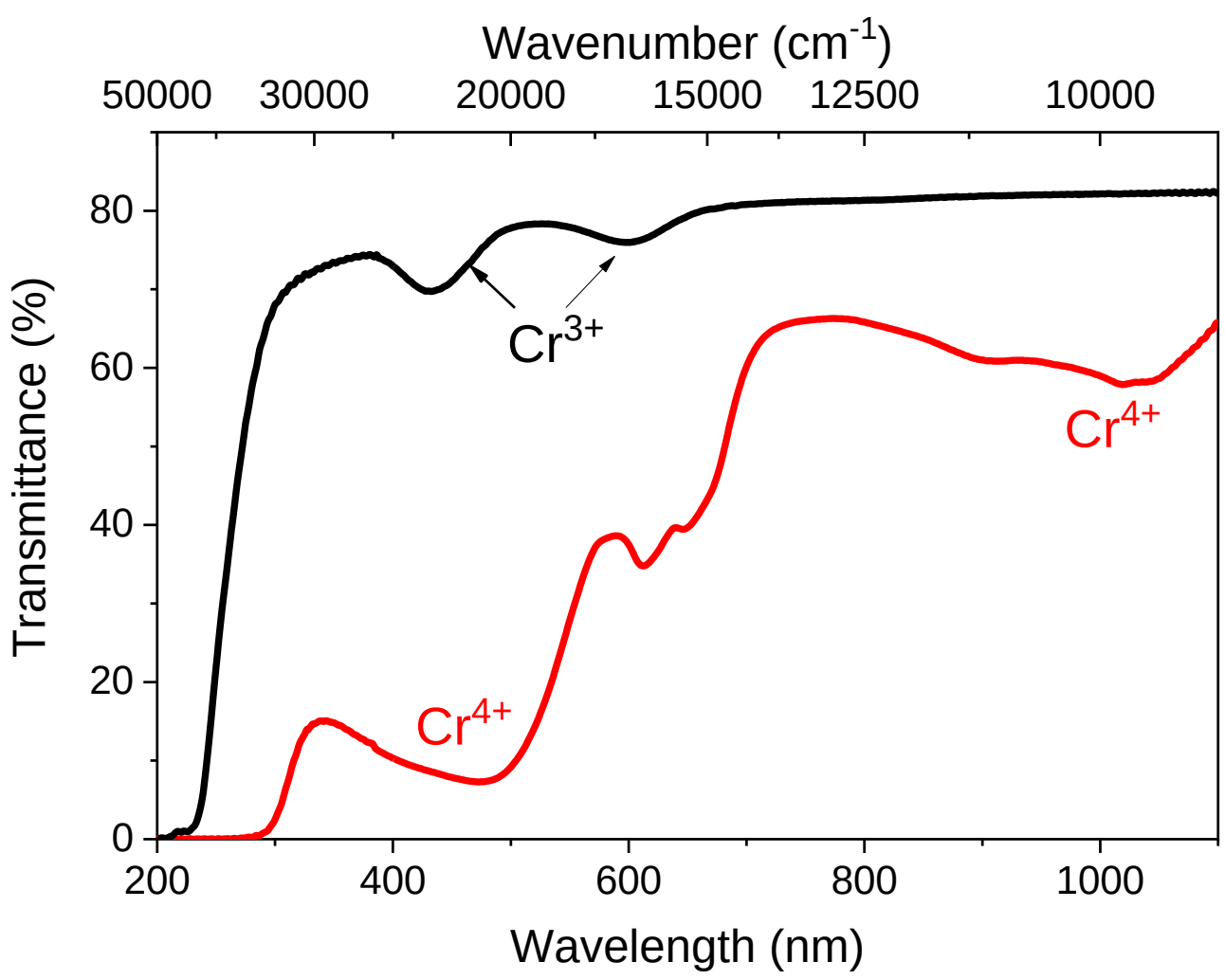

Fig. 4: Optical in-line transmission spectra of $Cr^{3+}$:YAG ceramics after vacuum sintering (black line) and air annealing (red line) at 300 K.

The transparent Cr:YAG ceramics were placed into a quartz vacuum tube and irradiated by a focused Nd:YAG laser beam with the emission wavelength of 1064 nm. The path of the laser beam was horizontal, running from left to right and penetrating the sample (Fig. 5). Here the thin unpolished side, the front polished side and the back polished side of the sample are designated as “edge”, “front” and “back” surfaces, respectively. In the course of LIWE experiments, it was found that LIWE occurs on the surfaces of the samples. Fig. 5(a) shows the case where the laser beam enters the sample at an acute angle to the “back” surface and exits the “front” surface. The white emission appears only at the entry (left) and the exit (right) points on of the “back” and “front” surfaces, respectively. The white emission wasn’t detected in the bulk of the samples.

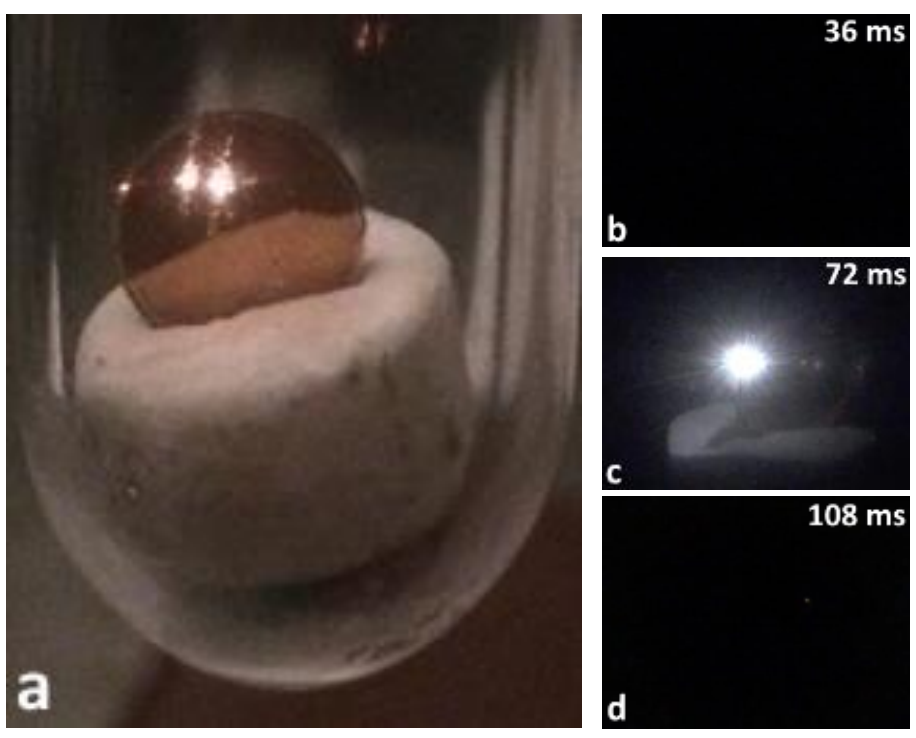

Fig. 5: a) The photo of LIWE under CW excitation of the transparent Cr:YAG ceramics. b-d) The set of the photographs taken every 36 ms when the sample is b) in front , c) under, and d) behind the focused laser beam.

The impact of the incident laser excitation on the LIWE during transition of the sample thorough the laser beam is shown in Fig. 5(b-d). The experiment was carried out by swinging the sample like a pendulum while the "front" and "back" surfaces remained parallel to the laser beam. The intense white emission was detected at the points where the sample crossed the laser beam (Fig. 5(c)). This emission disappeared when the sample wasn't under the laser beam (Fig. 5(b,d)). It was like a short flash of light as the sample passes through the laser beam (see Video 1). The color of the emission remains white even with this short excitation time. Given the short exposure time, it can be concluded that the LIWE can occur at relatively low host temperature.

LIWE spectra as a function of the laser power (maximum laser power was 3.4 W) are shown in Fig. 6. LIWE covers whole visible and near infrared region between 28000 $cm^{-1}$ (350 nm) and 10000 $cm^{-1}$ (1000 nm). These spectra can be fitted by two Gaussians with the maximums at 18500 $cm^{-1}$ (540 nm) and 15000 $cm^{-1}$ (660 nm) (Fig. S3). It should be noted that the FGS900S absorption filter was used to protect the CCD camera from the reflected laser beam. The correlation function was used to exclude the influence of the absorption filter on the emission spectra. Also the shape of obtained spectra has been modified by the sensitivity of the CCD camera.

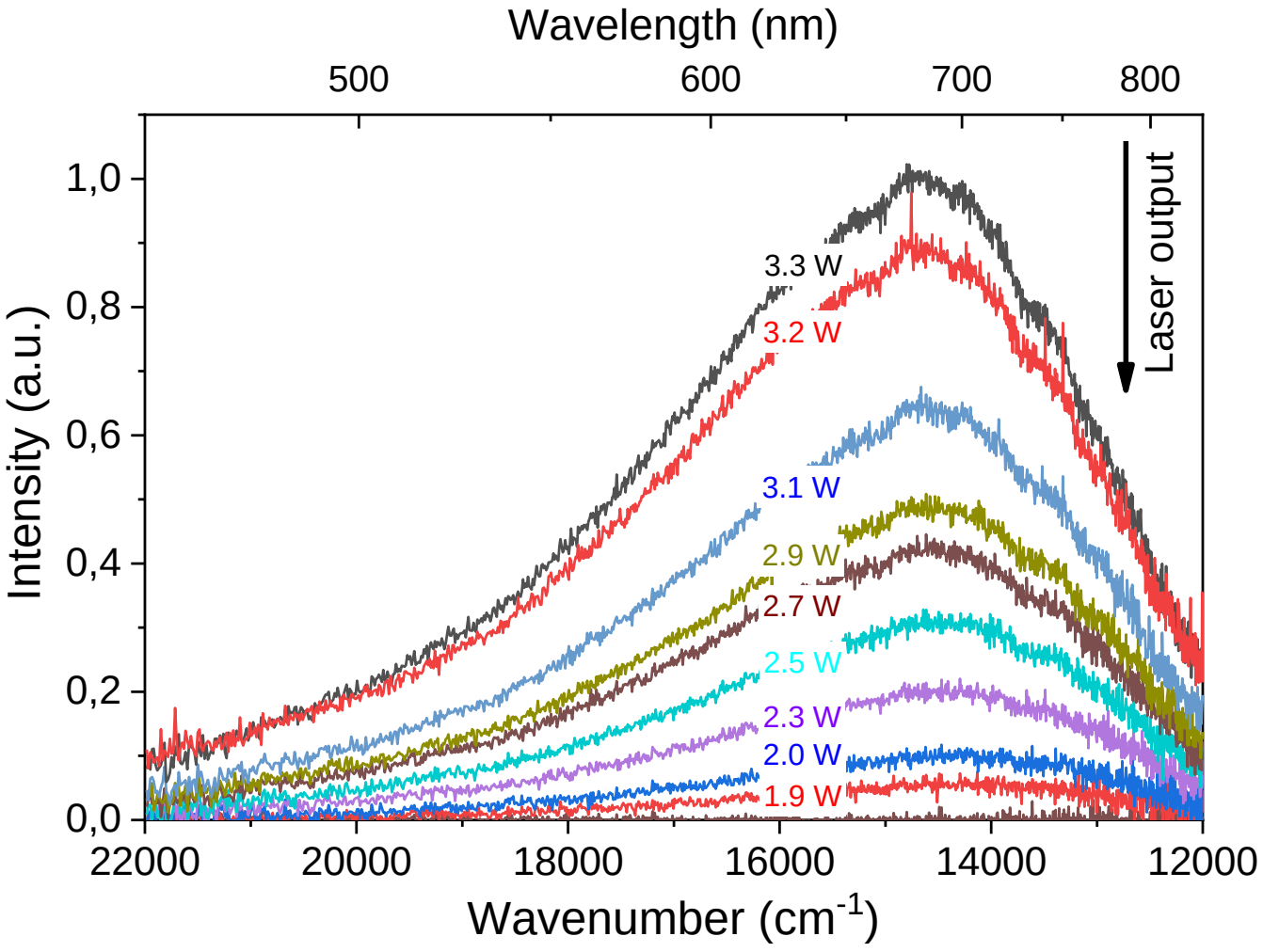

Fig. 6: LIWE spectra of a Cr:YAG ceramics upon excitation by focused 1064 nm laser.

The effect of pressure on LIWE of Cr:YAG ceramics at 1064 nm excitation is shown in Fig. 7. The pressure dependence of LIWE intensity has a threshold behaviour, while no LIWE is detected at atmospheric pressure. Decreasing pressure from 5 $10^{-5}$ to 0.1 mbar does not affect the emission intensity and some kind of saturation is observed. Further decrease in pressure below 0.1 mbar leads to decrease in LIWE by two orders of magnitude. It was shown earlier that this critical pressure linearly depends on the power of incident laser light. Wang et al. [19] suggested that the supralinear dependence of emission intensity against laser power can be predicted using a model where the powder in vacuum can only dissipate the incident laser energy as radiation. According to this model, the emission intensity follows an exponential decrease with pressure, and the LIWE intensity can be expressed as a function of the ambient pressure:

$$I_{LIWE} \propto I_o \cdot exp\left({}^{-P}/_{P_0}\right),$$

where $P_0$ is a critical value of pressure of the ambient atmosphere surrounding the host at which the intensity of white emission remains maximum. The parameter $P_0$ at 3.4 W (9.2 W/cm$^2$) excitation power is 0.6 mbar.

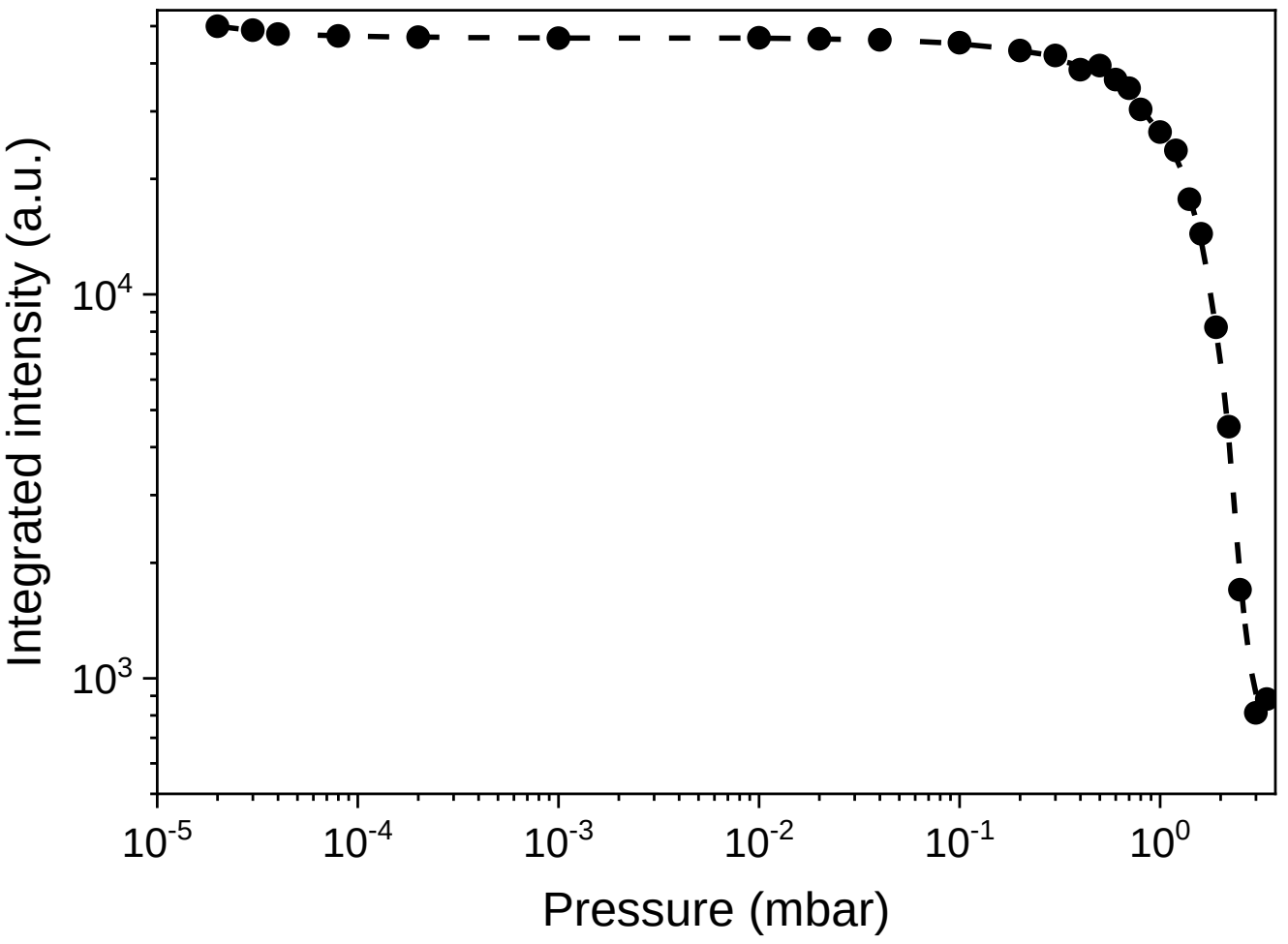

Fig. 7: The effect of the pressure inside the chamber on LIWE intensity upon excitation by focused 1064 laser (3.4 W).

As expected, the LIWE intensity showed a direct dependence of the emission intensity on the pumping power. LIWE intensity sufficiently increases with increasing pumping power, but only after exceeding the threshold value of the excitation power. Increasing the laser power from 0.1 to 0.7 W did not lead to appearance of LIWE and only when the value of 0.7 W was exceeded, a weak signal was detected.

It is well known that the number of photons required to populate an emission state can be determined from the slope of the power-law dependence plotted on a logarithmic scale using the following formula:

$$I_{em} \propto P^n$$

where $I_{em}$ is the emission intensity, P is the pump laser power, and n is the number of the required photons [7]. This relationship represents the n-photon absorption process and can be very roughly applied to LIWE. For Cr:YAG ceramics, the initial slope of the logarithmic plot of the emission intensity versus the incident laser power is 3.4±0.3 (Fig. 8). This indicates that 4 photons are involved into LIWE process corresponding to ~ 38000 $cm^{-1}$. The LIWE spectra consist of the broad band covering the region between 28000-10000 $cm^{-1}$. Therefore, this emission is characterized by a large Stokes shift of 10000-28000 $cm^{-1}$

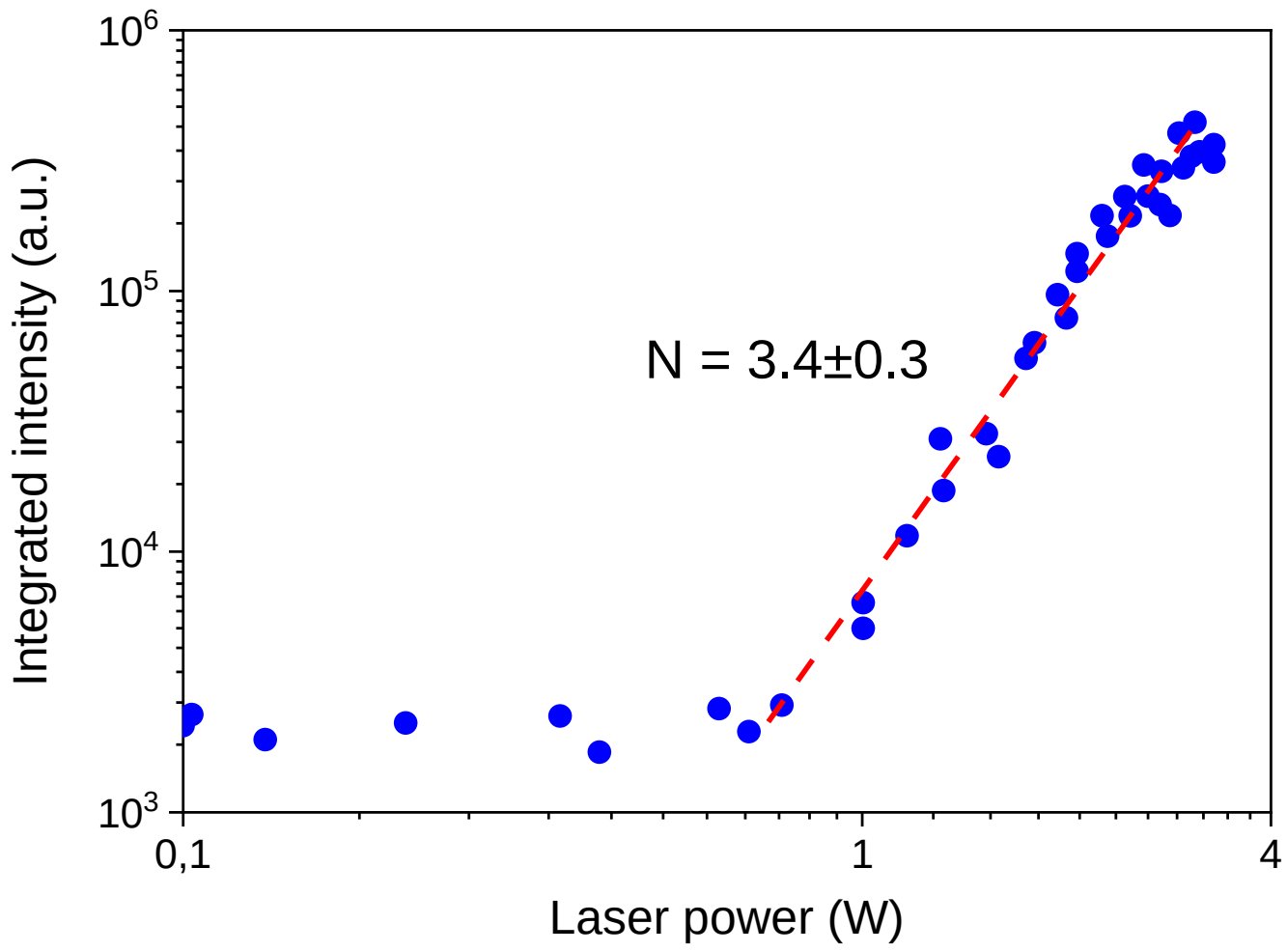

Fig. 8: The effect of laser power on LIWE intensity.

To characterize the colour of the emitted light, the International Commission on Illumination (CIE) coordinates (CIE1931, 2◦ Standard Observer) were determined for LIWE spectra at 1-3.1W for 1064 nm laser excitation (Fig. S4). Due to low emission intensity, the LIWE spectrum taken at 0.7 W power was not taken into account when calculating CIE colour coordinates. CIE coordinates are blue-shifted with increasing laser power.

The time dependence of LIWE intensity was measured (1064 nm laser excitation, 3.4 W) (Fig. S5). The LIWE intensity increased with an increase in the excitation time and after a certain time reached a maximum with a subsequent decrease. The quantum efficiency of $^3B_2(^3A_2)$-$^3B_1(^3T_2)$ transitions in $Cr^{4+}$ ions is low (only 15%) at room temperature [20].

Therefore, the rise of host temperature with increase of the time of laser excitation was expected. The increase in LIWE intensity is probably associated with an increase in the host temperature. This assumption is consistent with [1], where LIWE intensity increase in more than 10 times at host temperature rise from 10 K to 300 K was shown.

The host temperature of the Cr:YAG ceramics during LIWE was estimated from $Cr^{3+}$ emission spectra at simultaneous excitation by 404 nm (0.3 W) and 1064 nm (3.4 W) lasers. The 1064 nm laser was focused on the edge, so the laser light passed through the entire sample and exited from the opposite side. The blue laser beam was perpendicular to the infrared one and illuminated a small volume at the edge. Emission spectra were measured every 0.2 s. The 404 nm laser was continuously on, while the 1064 nm laser was switched on at 0 s. Fig. 9 shows the time evolution of LIWE spectra under double excitation. This experiment was performed on a 0.3 mm thick sample to ensure that the same area was excited. Switching on the laser excitation leads to strong red emission of $Cr^{3+}$ ions (Fig. 9(a)). The $Cr^{3+}$ emission intensity decreased with time in contrast to LIWE which increased in the first 2.2 s. Further time increase leads to decrease in the LIWE intensity (Fig. 9(e)) observed probably due to an increase in the host temperature. The host temperature can be estimated from $Cr^{3+}$ emission spectra. Fig. 10 shows the temperature dependence of $Cr^{3+}$ luminescence measured using the same equipment. For comparison, Fig. S6 shows the emission spectra of a Cr:YAG ceramic excited by 433 nm lamp at 300 K measured using FLS980 Fluorescence Spectrometer.

The emission spectra of $Cr^{3+}$ ions in YAG lattice consist of the broadband emission ascribed to the $^4T_{2g}\rightarrow{}^4A_{2g}$ fluorescence with a maximum at 707 nm, the sharp and narrow R-lines corresponding to $^2E_g\rightarrow{}^4A_{2g}$ transitions at 688 nm, the diffused band centred at 675 nm (anti-Stokes vibronic sidebands), and two sharp diffused bands at 703 nm and 725 nm corresponding to the vibronic sidebands assisting the fluorescence transition [21,22]. The shape of $Cr^{3+}$ emission spectra strongly depends on the host temperature. Fig. 10 shows the temperature dependence of $Cr^{3+}$ emission spectra upon 405 nm excitation in the range of 30–600°C. The sharp R-lines were observed at room temperature, while temperature increase leads to decrease in the intensity of the R-lines disappearing completely at 400°C [23]. Therefore, the host temperature during LIWE can be estimated by monitoring intensity of R-lines.

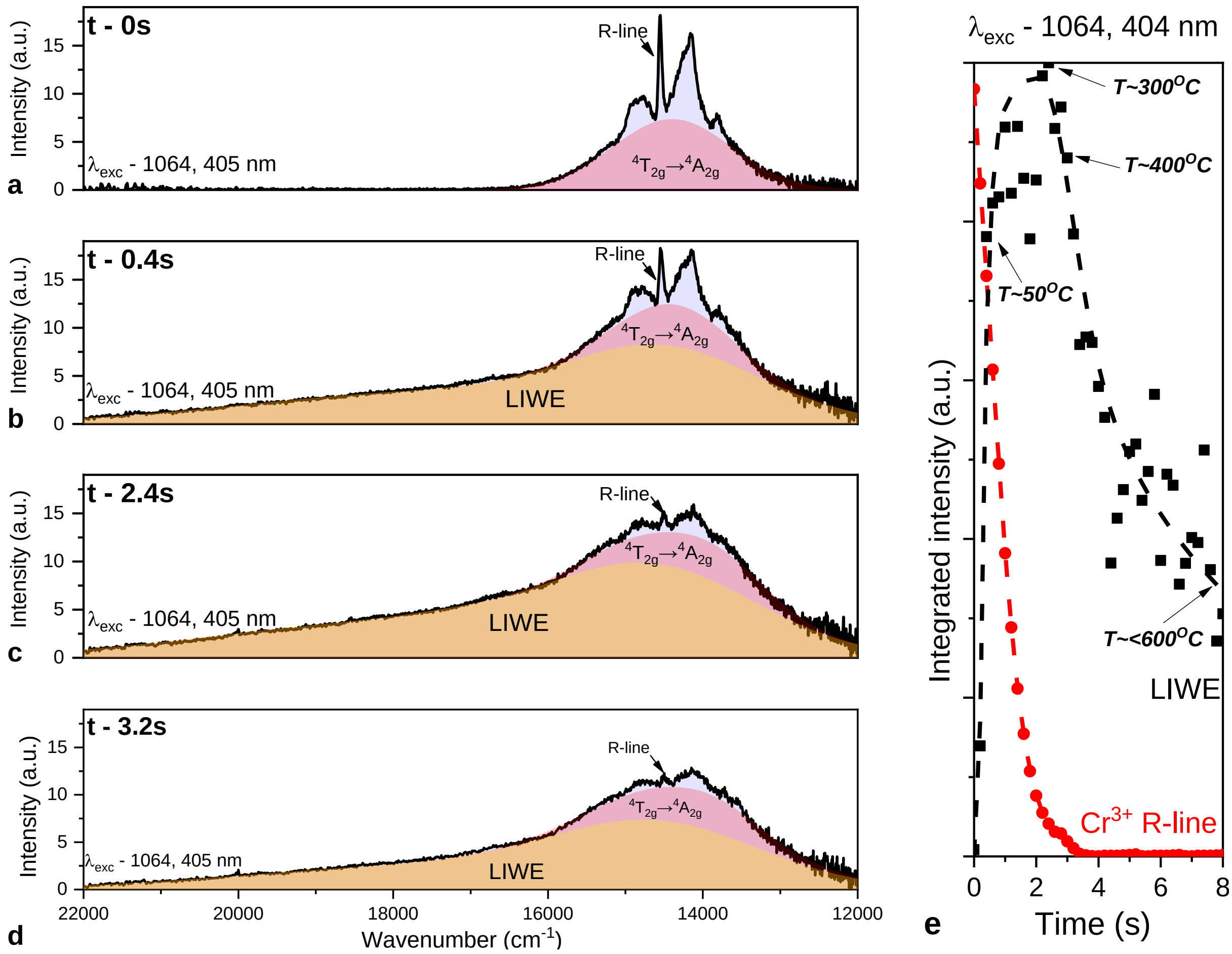

Fig. 9: The time evolution of a-d) LIWE and $Cr^{3+}$ spectra and e) integrated intensity of LIWE (black point) and $Cr^{3+}$ R-lines (red point) from a Cr:YAG ceramics simultaneously excited by 1064 nm (3.4 W) and 405 nm (0.4 W) lasers.

Fig. 9(e) shows the temporal evolution of the LIWE and the intensities of $Cr^{3+}$ R-lines. The intensity of R-lines decreased with time and these lines disappeared after 3.4s. So, the host temperature increases over time and exceeds 500°C after 3.4s. The LIWE intensity increased with time reaching maximum at 2.4s followed by decrease in the emission intensity. Due to the strong overlap with the $Cr^{3+}$ emission, the LIWE intensity was measured in the range from 22000 cm$^{-1}$ to 17000 cm$^{-1}$. The LIWE intensity reached 75% in just 0.4s, where 100% is the maximum value of LIWE intensity found during the experiment (Fig. 9(b)). At the same time, the intensity of R-lines was 73% of the value observed at room temperature (25°C) which means that host temperature was about ~50°C at 0.4s. The maximum intensity of LIWE was found at 2.4s while intensity of R-lines was ~6% (Fig. 9(c)). This means that the host temperature was ~290°C at 2.4s. The LIWE intensity drops below 75% after 3.2s, while the intensity of R-lines was 1.5% corresponding to ~400°C (Fig. 9(d)). Even after 8s, the weak signal of R-lines, $^2E_g \rightarrow ^4A_{2g}$ vibronic sidebands and $^4T_{2g} \rightarrow ^4A_{2g}$ broadband emission can be recognized in the emission spectra indicating that the host temperature remains below 600°C. It should be noted that the 1064 nm laser cross section was up to 0.18 mm, with the sample thickness of 0.3 mm. This means that 60% of the thickness was irradiated by both lasers, and the remaining 40% was irradiated by only 404 nm laser. Due to the high thermal conductivity of YAG ceramics, one cannot expect a high temperature gradient.

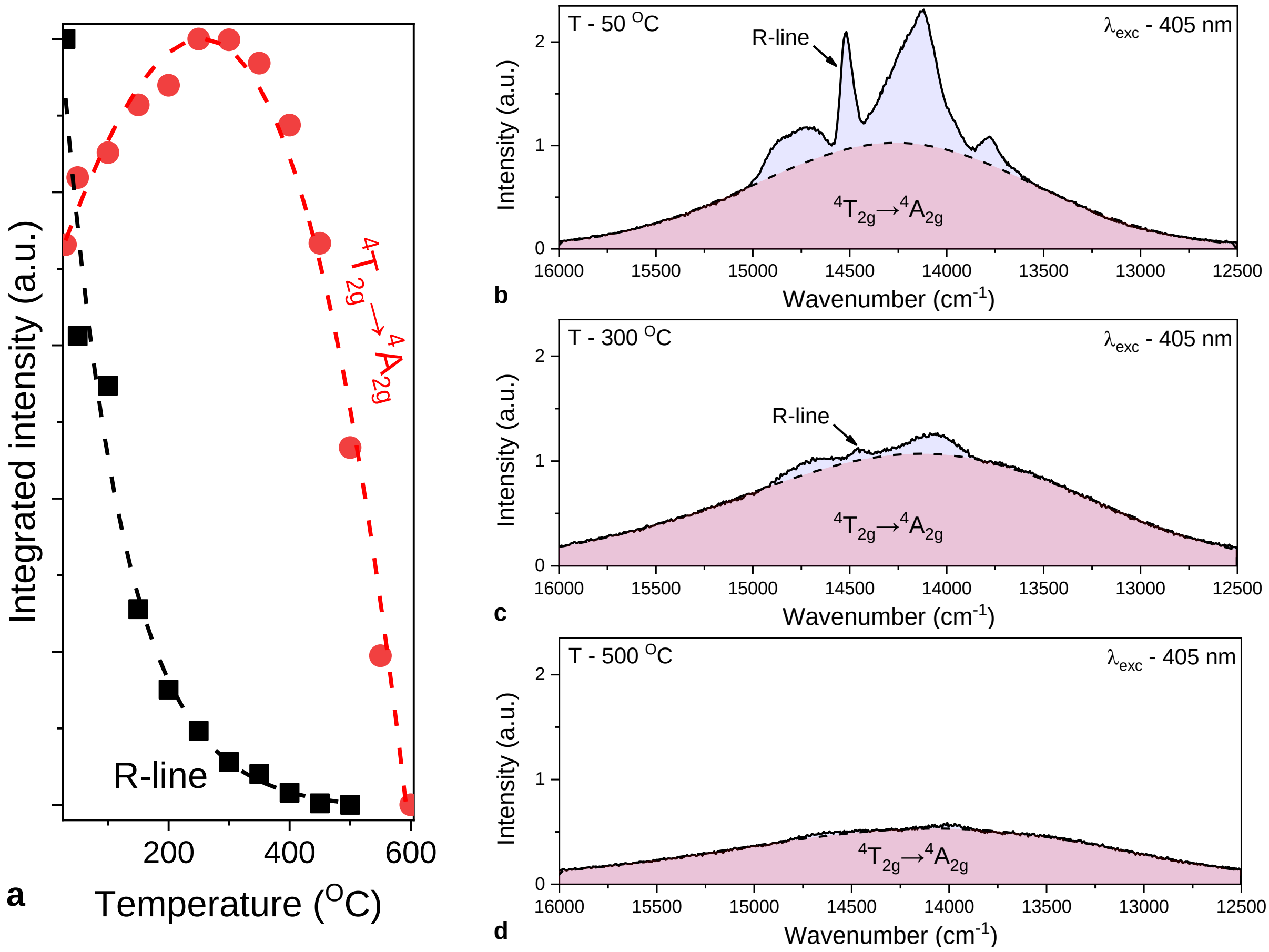

Fig. 10 a) Temperature dependence of the integrated intensity of the R-lines, and the $^4T_{2g} \rightarrow ^4A_{2g}$ emission. Luminescence spectra of a Cr:YAG ceramics under 405 nm laser excitation at b) 50°C, c) 300°C and d) 500°C, the red and blue areas correspond to the $^4T_{2g} \rightarrow ^4A_{2g}$ and $^2E_g \rightarrow ^4A_{2g}$ electron transitions, respectively.

The host temperature is an important parameter for understanding LIWE, since some of the authors suggest that LIWE arose as a result of thermal emission of the host under focused laser beam. The energy of laser beam can be accumulated in the host due to its low thermal conductivity. In our previous studies, the host temperature during LIWE was measured by luminescence thermometry [6] and thermovision camera [7]. The host temperature during LIWE was 800°C for $LiYbF_4$ [6] and 380°C for NdAP [7] nanocrystals under focused 980 nm laser (2.1 W, ~$5.5·10^3$ W/cm$^2$). Our results suggest that the temperature of Cr:YAG ceramics during LIWE was below 600°C under focused 1064 nm laser (3.0 W, ~$9.2·10^3$ W/cm$^2$) that is consistent with the literature data. The highest LIWE intensity was found in the range from 50°C to 400°C. We assume that decrease in the LIWE intensity after a certain excitation time was caused by heating of the host above 300°C in case of transparent Cr:YAG ceramics. Consequently, the host temperature is too low for the thermal origin of LIWE.

## Discussion

We suppose that LIWE phenomena can be described by IVCT mechanism. Earlier, this model was proposed for Yb-doped YAG materials [24]. IVCT mechanism in $Yb^{3+}/Yb^{2+}$ pairs is described in details in [24]. This model suggests the following steps: the first stage associated with the mechanism leading to promotion of the electrons to the conduction band (CB) and the second one which is responsible for broadband emission in the visible range [25]. Promotion to CB can be realized by multiphoton absorption and multiphoton ionization or cooperative energy transfer from two exited tetrahedral $Cr^{4+}$ ions to chromium mixed valence pair. The second step is the electron transfer from donor to acceptor and, finally, recombination with acceptor from chromium mixed valence pair (Fig. 11). As trivalent and tetravalent chromium ions are most commonly found in the Cr:YAG ceramics, we assume that the IVCT process occurs in a $Cr^{3+}/Cr^{4+}$ mixed valence pair. However, the possible combinations of mixed valence pairs also can include $Cr^{2+}$, $Cr^{5+}$, $Cr^{6+}$ ions or even some structural defects [26,27].

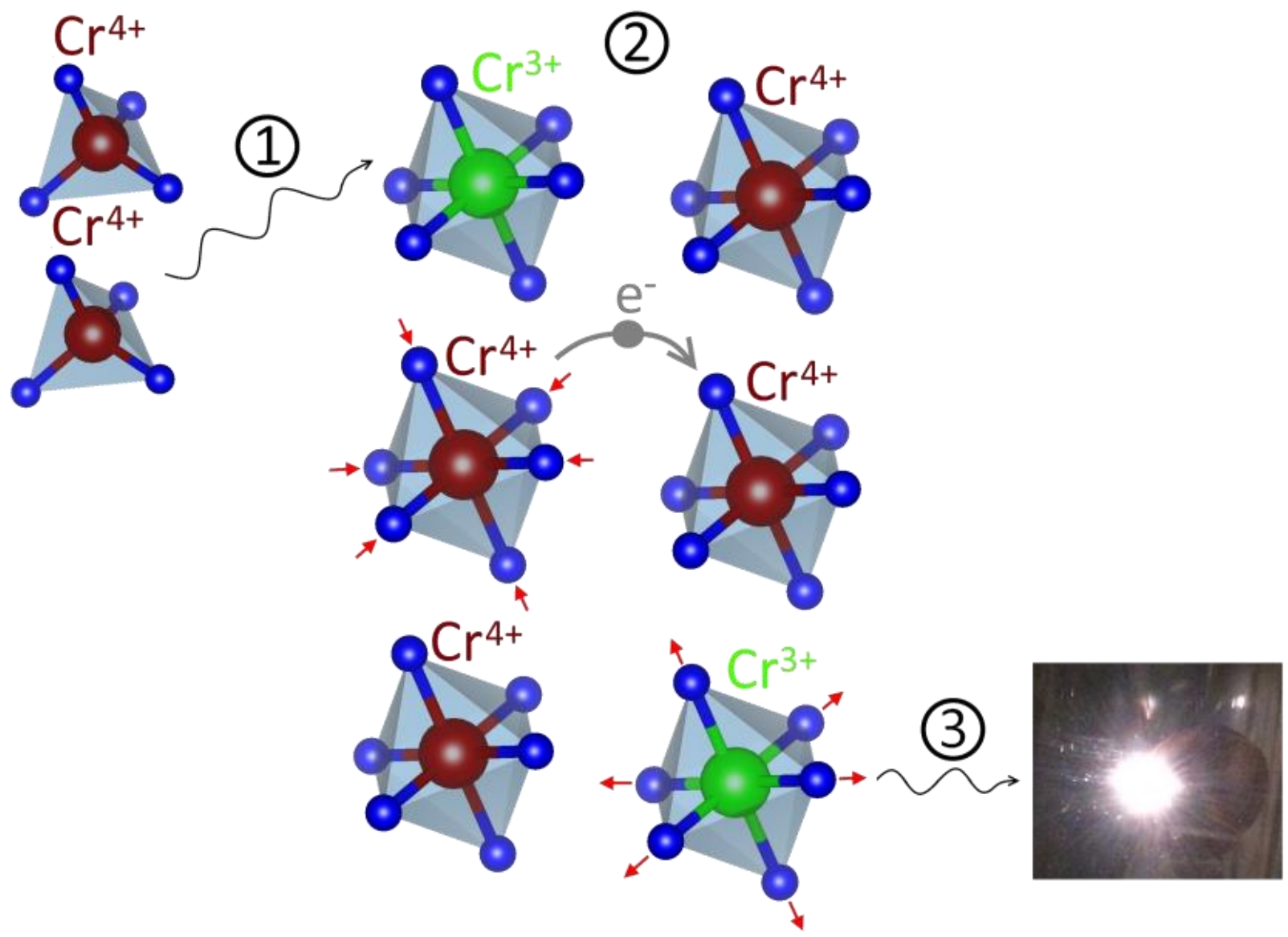

Fig. 11: General scheme of LIWE in Cr:YAG ceramics. The numbers indicate the steps of the process: 1) cooperative energy transfer from excited tetrahedral $Cr^{4+}$ ions to $Cr^{3+}/Cr^{4+}$ mixed valence pair; 2) IVCT in $Cr^{3+}/Cr^{4+}$ pair and 3) LIWE. The red arrow indicate reducing/increasing the Cr-O distance

Two different mechanism of the promotion of electrons to CB can be proposed. The first one, which is based on the tunnelling effect, for which the ionization of atoms take places under high-power radiation with frequency lower than the ionization potential, was proposed by Keldysh [28]. The indirect transition of an electron from the ground state of an atom to a free state is based on multi-photon absorption process. The second is the cooperative sensitization upconversion. IVCT occurs in $Cr^{3+}/Cr^{4+}$ mixed valence pair if both ions occupy octahedral sites (see discussion below). Since both octahedral $Cr^{3+}$ and $Cr^{4+}$ ions have no absorption bands in the near-infrared region, the mechanism leading to promoting the electrons to the conduction band can be based on the energy transfer from tetrahedral $Cr^{4+}$ ions to $Cr^{3+}/Cr^{4+}$ mixed valence pair. Schematic potential surface diagram for an octahedral $Cr^{3+}$, an octahedral $Cr^{4+}$ and tetrahedral $Cr^{4+}$ complexes depicting its relationship with electronic configurations and with absorption spectra is shown in Fig. 12. Energy transfer between ions takes place when the sensitizer ($Cr^{4+}$) is in one of its excited states, while the activator ($Cr^{3+}/Cr^{4+}$ ions pair) is in its ground state; then, the energy obtained by the sensitizer is transferred to the activator [29]. The energy transfer can occur during cooperative sensitization, while the energy obtained by two excited tetrahedral $Cr^{4+}$ ions transfers to chromium mixed valence pair at once [29]. Absorption of 1064 nm radiation leads to electronic transition from $^3E(^3T_2)$ and $^3A_2(^3B_1)$ levels. The excited level decays due to the electron-phonon interaction to the $^3B_2(^3T_2)$ level, excited state absorption occurs between the $^3B_2(^3T_2)$ and $^3E(^3T_1)$ levels [20]. Consequently, two tetrahedral $Cr^{4+}$ ions can accumulate the energy equal to four 1064 nm photons involved in LIWE process.

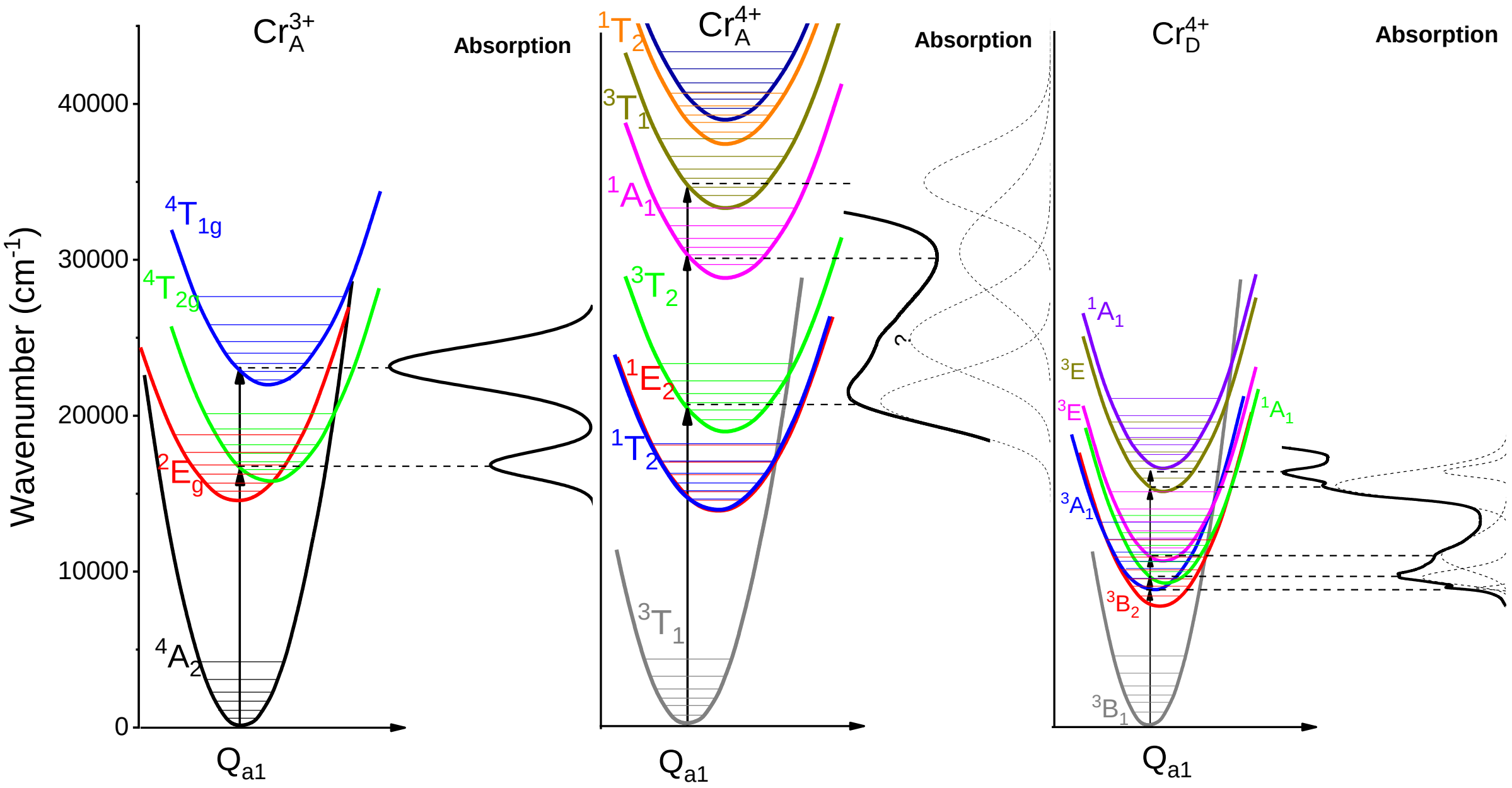

Fig. 12 Schematic potential surface diagram for an octahedral $Cr^{3+}$ ($Cr_A^{3+}$) [23], an octahedral $Cr^{4+}$ ($Cr_A^{4+}$) [30] and tetrahedral $Cr^{4+}$ ($Cr_D^{4+}$) [31] complexes depicting its relationship to electronic configurations and absorption spectra adapted from [10].

The second stage, associated with broadband emission in the visible range, can be described by IVCT mechanism. The large Stokes shift and broad emission bands were associated with change in the Cr-O distance due to the difference in ionic radii of $Cr^{3+}$ and $Cr^{4+}$ ions. According to IVCT model [24], the $Cr^{3+}$ and $Cr^{4+}$ were considered not as a single ions but as a defect centres including distortions in the first coordination shells due to the difference between ionic radii of dopant and host ions [24]. Let us label the two $Cr^{3+}$ and $Cr^{4+}$ ions as $Cr_L^{3+}$ and $Cr_R^{4+}$. The IVCT in $Cr_L^{3+}/Cr_R^{4+}$ pair leads to electron transfer from $Cr_L^{3+}$ to $Cr_R^{4+}$ in result forming $Cr_L^{4+}/Cr_R^{3+}$ pair (see Fig. 11, process 2). The valence change of $Cr^{3+}$ to $Cr^{4+}$ ion caused by the difference in ionic radii of $Cr^{3+}$ and $Cr^{4+}$ ions, leads to decrease of the Cr-O distance and vice versa (the red arrows in Fig. 11). These transformations require high reorganization energies causing large Stokes shifts. In this way, IVCT transitions lead to formation of structurally stressed states, and so, to very broad emission bands (Fig. 6).

The IVCT process in $Cr^{3+}/Cr^{4+}$ mixed valence pair can occur when both ions occupy octahedral sites. The YAG formula can be written as [C3][A2][D3]O12, where C, A, and D denote cations in dodecahedral, octahedral, and tetrahedral positions, respectively [11]. An interesting feature of $Cr^{4+}$ ions in YAG matrix is the ability to occupy both octahedral and tetrahedral crystallographic positions. $Cr^{3+}$ ions occupy octahedral site, while $Cr^{4+}$ ions occupy both octahedral and tetrahedral sites. After vacuum sintering, all chromium is in the trivalent state and occupies octahedral sites [10]. During air annealing, octahedral $Cr^{3+}$ ions lose electrons leading to formation of octahedral $Cr^{4+}$ ions [18]. Tetrahedral $Cr^{4+}$ ions appear due to an intra-lattice cation exchange between octahedral $Cr^{4+}$ ion and tetrahedral $Al^{3+}$ ion leading to the formation of octahedral $Al^{3+}$ ion and tetrahedral $Cr^{4+}$ ion [27]. Intra-lattice cation exchange is a thermally activated process and occurs at temperatures above 800°C [31], that is lower than the host temperature during LIWE. So, an intervalence charge transfer in $Cr^{3+}/Cr^{4+}$ mixed valence pair can occur when both ions occupy the octahedral sites.

Based on IVCT models, the impact of LIWE on the performance of $Cr^{4+}$:YAG-based laser can be predicted. We suppose that LIWE and, as result, the laser efficiency can be improved by changing the Cr-O distance by varying the crystalline matrix. The energies of the electron levels of the $Cr^{3+}/Cr^{4+}$ ionic configuration vary with the displacement of the oxygen atoms around these ions [24]. So, the change in distance between the $Cr^{4+}$ and $O^{2-}$ ions can be critical to the efficiency of LIWE in $Cr^{4+}$:YAG materials. This distance can be modified by choosing a different crystal matrix, for example GGG or/and using different divalent dopant. The influence of divalent dopant can be significant because $Cr^{4+}$ ions exist as $[Me^{2+}...Cr^{4+}]$ complex, where $Me^{2+}$-$Ca^{2+}$ or/and $Mg^{2+}$ [32,33]. Therefore, the type of divalent dopant and/or matrix can be critical to the LIWE efficiency and, as a result, can affect the laser properties of $Cr^{4+}$-doped materials. This hypothesis is consistent with the published data on laser characteristics of $Cr^{4+}$-doped garnet. The type of divalent dopants and/or hosts affect the laser efficiency of $Cr^{4+},Ca^{2+}$:YAG, $Cr^{4+},Mg^{2+}$:YAG and $Cr^{4+},Ca^{2+}$:GGG Q-switchers [34].

**Conclusion**

We have found that transparent Cr:YAG ceramics is able to generate bright LIWE at excitation power above a certain threshold. LIWE covers whole visible and near infrared region between 28000 $cm^{-1}$ (350 nm) and 10000 $cm^{-1}$ (1000 nm). It was shown that LIWE intensity increases with increasing excitation density. There are four photons involved in the LIWE process. LIWE intensity increased with an increase in the excitation time and reached a maximum with a subsequent decrease. An increase in LIWE intensity is probably caused by an increase in host temperature. The host temperature during LIWE was estimated from $Cr^{3+}$ emission. It was shown that the maximum LIWE intensity was in the host temperature range from 50 to 400°C with a maximum at 300°C. The host temperature during LIWE did not reach 600°C. The IVCT model was used to explain the LIWE phenomena.

**Acknowledgement**

This work was supported by Polish National Science Centre, grant: PRELUDIUM-18 2019/35/N/ST3/01018.